\documentstyle{article}
\def\name#1{\hbox to 3in{#1\hfil}}

\newcommand{\bP}{\bf P}
\newcommand{\bZ}{\bf Z}
\newcommand{\bC}{\bf C}
\def\O{{\cal O}}
\begin{document}
\title{\Large A Non-Perturbative Superpotential With $E_8-$Symmetry
      }
        \author{Ron Donagi\\
Department of Mathematics, University of Pennsylvania\\ \\
Antonella Grassi\\
Department of Mathematics, University of Pennsylvania\\ \\
Edward Witten
\\
School of Natural Science, 
Institute for Advanced Study, Olden Lane, Princeton, New Jersey
}
\maketitle
\bigskip

\medskip
\large

\noindent

{\bf Abstract}

We compute the non-perturbative superpotential in $F$-theory
compactification to four dimensions on a complex three-fold
${\bf P}^1\times S$, where $S$ is a rational elliptic surface.
In contrast to examples considered previously, the superpotential
in this case has interesting modular properties; it is essentially
an $E_8$ theta function.

\S 1. \bf Introduction\rm.

\medskip

Compactifications of $M$-theory to 3 dimensions, and of $F$-theory to
4 dimensions, on a 4-complex-dimensional Calabi-Yau manifold $X$
were studied in [1].  There it was shown that the superpotential for
these theories is generated entirely by instantons obtained by
wrapping a 5-brane over a complex divisor $D\subset X$ which must
further
satisfy $\chi(D, \O _D)=1$.  In all cases considered in [1], there were a
finite number of such divisors $D$, yielding either a vanishing
superpotential or a fairly simple non-zero one, in contrast to
speculations [2,3] according to which non-perturbative superpotentials
might show modular behavior.

     Here we study a particular compactification based on a
Calabi-Yau fourfold $X$ which contains an infinite collection of
contributing divisors.  The resulting superpotential turns out to
be essentially the theta function of an $E_8$ lattice, and thus
in particular has modular properties that presumably are related
to an otherwise not apparent duality symmetry of this compactification.
After computing the superpotential, we exhibit a vacuum state
with unbroken supersymmetry.

  At the moment, we do not
know how the $E_8$ symmetry of the superpotential we obtain is
 related to the $E_8\times
E_8$ gauge symmetry of the heterotic string.
The calculation we perform
also has an application, which we explain at the
end of the paper, to the counting of BPS states in a certain
compactification to six dimensions; in that case, the relation
to $E_8$ is better understood via string dualities.

\medskip

\noindent

\S 2. \bf Review of non-perturbative superpotentials \rm.

\medskip

     Let $X$ be a Calabi-Yau fourfold.  The superpotential for
$M$-theory compactified on $X$ is shown in [1] to come from 5-branes
wrapped
over complex divisors $D\subset X$.  An anomaly computation
shows
that not all such $D$ can make a non-zero contribution to the
superpotential; a necessary condition is that
$$\chi(D,\O_D):=h_0-h_1+h_2-h_3=1,$$
where $h_i$ is the dimension of the space of holomorphic $i$-forms
on $D$.  An example of this occurs when $h_1=h_2=h_3=0$.  In
this situation (and only then) it is known that the contribution of
$D$ to the superpotential is non-zero.

 In view of $F$-theory, a particularly interesting
 case arises when $X$ is elliptically
fibered,
i.e. there is a map $\pi :X\rightarrow B$ whose generic
fibers are elliptic
curves.  (It is usually also required that $\pi$ have a section.)  In
this case [4], $M$-theory compactified on
$X$ has a limit which is equivalent to a Type IIB superstring
compactified on the base $B:\; \; M$-theory on $X$ is equivalent to Type
IIB on $B\times S^1$. If $\epsilon$
is the area of the
elliptic fibers of $\pi$, then the length of the $S^1$ scales
like $\epsilon^{-1}$, so the limit as $\epsilon\to 0$
gives the IIB superstring compactified on $B$.

     The divisors $D\subset X$ can be classified as horizontal (if
$\pi (D)=B$) or vertical (if $\pi (D)$ is a divisor $C\subset B$).
An order-of-magnitude argument in [1] shows that, while both
horizontal and vertical divisors with $\chi(\O_D)=1$
 can contribute to
the superpotential in the IIB theory compactified on $B\times S^1$,
only contributions from the vertical divisors survive the
$\epsilon \rightarrow 0$ limit.

\def\Ch{{\rm Ch}}
\def\Td{{\rm Td}}
\def\O{{\cal O}}
     We therefore need to find those smooth divisors $C\subset B$
such
that $\pi^{-1}(C)$, or more generally some component $D$ of $\pi^{-
1}(C)$, satisfies $\chi(\O_D)=1$.  This search is simplified by the following
argument.\footnote{The following Riemann-Roch computation was
carried out independently by A. Klemm.}
We assume for simplicity that $\pi^{-1}(C)$ is always
irreducible.  This will be the case, for example, if all the fibers
of $\pi$ happen to be irreducible.

\medskip

     We recall [5, Appendix 1], that the Hirzebruch Riemann-Roch
theorem for a line bundle $\O_X(D)$ on $X$ says that
$$\chi(\O(D))=(\Ch(\O(D))\cdot \Td(X))_{top},$$
where
$$\Td(X)=1+\frac{c_1}{2}+\frac{c^2_1+c_2}{12}+\frac{c_1c_2}{24}-
\frac{c^4_1-4c^2_1c_2-3c^2_2-c_1c_3+c_4}{720} + ...$$
and
$$\Ch(\O(D))=1+D+\frac{D^2}{2}+\frac{D^3}{6}+\frac{D^4}{24}+....$$
When $X$ is 4-dimensional, this gives
$$\chi(\O(D))=\frac{D^4}{24}+\frac{D^2 c_2}{24}+\frac{c^2_2}{240}-
\frac{c_4}{720}+(\mbox{multiples of }c_1),$$
and the terms involving $c_1$ will vanish when $X$ is Calabi-Yau.

Using the short exact sequence
$$0\rightarrow \O_X(-D)\rightarrow \O_X\rightarrow \O_D\rightarrow 0$$
we find that
$$\chi(\O_D)=\chi(\O_X)-\chi(\O_X(-D))=
- - -\frac{D^4+D^2c_2}{24}.$$
In the case of interest,
since $D$ is a pullback $\pi^{-1}(C)$, its four-fold self-intersection
vanishes,
$$(D^4)_X=\pi^{-1}(C^4)_B=0,$$
so we are left with $\chi(\O_D)=-\frac{1}{24} {D^2 c_2}$.

     A divisor $D\subset X$ is called \it {nef}  \rm (formerly,
the term was ``numerically effective") if its intersection number
with every effective curve $A$ in $X$ is $\geq0$
 (see, for example [6]).
 This holds if
$D$ is ample, but also for $D=0$.  More generally, if
$f:X\rightarrow Y$ is a morphism and $C$ is nef on $Y$, then
$D:=f^{-1}(C)$ is nef on $X$.
If  $D\cdot A < 0$, then $D$ must contain $A$,
and this situation must persist whenever $D$ or $A$ are deformed.

The well-known second Chern-class inequality says that for a nef
divisor $D$ in an $n$-dimensional CY manifold $X$,
$$D^{n-2}\cdot c_{2}\geq 0.$$
For ample $D$, this is due to Yau [7]. The general nef case 
follows from this since a nef divisor is a limit of ample ones. 
An algebraic proof of the nef case was given by Miyaoka [8, Coro. 6].
(There is also a more general result for arbitrary $X$, with additional
terms involving $c_1$.) 

 Combining this with our previous
calculation, we conclude that for a vertical divisor $D=\pi^{-
1}(C)$ in our elliptically fibered Calabi-Yau $X\rightarrow B$ to
have $\chi(\O_D)>0$, both $D$ and $C$ must be non-nef.  A number of
illustrations of this can be found in [1], where the base $B$
contained a finite number of exceptional divisors $C$ (or none).

\medskip

\noindent

\S 3. \bf  The Calabi-Yau fourfold. \rm

\medskip

     We start with a del Pezzo surface $S$ obtained from $\bP^2$ by
blowing up $9$ points $p_1, \cdots, p_9$.
  We assume that no $3$ of these are colinear,
and no $6$ lie on a conic.  If the points are in general position,
there will be a unique plane cubic curve through them.  We will
also be interested in the special case where the 9 points are the
complete intersection of two transversal cubics, $A=0$ and $B=0$,
so there is a whole pencil of cubics $\lambda A+\mu B=0,\;
(\lambda,
\mu)\in \bP ^1$, through the points.  In this case our surface is a
hypersurface of bidegree (3,1) in $\bP^2\times \bP ^1$, given by the
vanishing of the polynomial
$$f:=\lambda A+\mu B.$$
We let $p:S\rightarrow \bP^1$ denote the elliptic fibration in this
case.  A detailed study of del Pezzo surfaces can be found in [9],
and we will return to them in the next section.

     For the base of the elliptic fibration $\pi: X\rightarrow B$ we
take $B:=S\times \bP ^1\subset \bP^2\times \bP ^1 \times \bP ^1$.  We
take $X$ to be given over $B$ by a Weierstrass equation
$$g := y^2z-(x^3+\alpha xz^2+\beta z^3) = 0 $$
where $\alpha,\beta$ are sections of appropriate bundles on $B$.
In the special case above, we can write $X$ as the complete
intersection in $\bP^2 \times \bP ^1 \times \bP ^1 \times \bP^2$ of the
polynomials $f$, of multidegree $(3,1,0,0)$,
and $ g$, of multidegree
$(0,1,2,3)$.  (These degrees assure, by adjunction, that $X$ is
CY.)  An easy Bertini argument shows that for generic
values our $X$ will be non-singular.

     In the special case we note that $X$ has a second elliptic
fibration $\tilde{\pi}: X\rightarrow \tilde{B}$ where
$\tilde{B}$ is the hypersurface $g=0$ in $\bP ^1 \times \bP ^1 \times
\bP^2$.  It in turn has a $K3$-fibration $\kappa$
over (the first) $\bP ^1$: the fiber over $(\lambda ,\mu)\in \bP ^1$ is
a $K3$ surface $K_{(\lambda,\mu)}\subset \bP ^1 \times
\bP^2$, which itself has an elliptic fibration over (the second)
$\bP ^1$, as well as a realization as a double cover of $\bP^2$.  The
base $\tilde{B}$ is rational: it can be identified as the blowup of
$\bP ^1 \times \bP^2$ along the curve
 $ \Gamma \subset \bP ^1 \times \bP^2$,
of genus 28, which is the base locus of the pencil of $K3$
surfaces $K_{(\lambda,\mu)}$.  The fibration $\tilde{\pi}$ is
easy to understand:  it arises in a fiber-product diagram,

$${ \matrix{ & X & \longrightarrow  & S \cr 
        {\tilde \pi}& \downarrow &      &\downarrow  \cr 
        & {\tilde B} &\longrightarrow & {\bf P^1} \cr
        & & \kappa &}}$$

In other words, the fiber $\tilde{\pi}^{-1}(b)$ for $b\in
\tilde{B}$ depends only on $\kappa (b)\in \bP ^1$, and is nothing but
the elliptic curve $\lambda A+\mu B=0$.

\medskip

\noindent

\S 4. \bf  The del Pezzo surface. \rm

\medskip

We want to find all non-nef, smooth irreducible divisors $C$ in
$B= S\times \bP ^1$. The ${\bP}GL(2)$-action on $\bP ^1$ lifts to $B$,
showing that a non-nef divisor must be vertical, $C=E\times \bP ^1$ for
some (necessarily non-nef) divisor  $C$ in $S$.
In fact,  if $C$ is not vertical
and $A$ is any irreducible, effective curve in $B$,
then either $A$ is
a $\bP ^1$ fiber (in which case $A\cdot C \geq 0$, since $A$
moves), or $C$ can be moved by $PGL(2)$ so as not to contain $A$,
so again $A\cdot C\geq 0$.

So we are looking for the non-nef irreducible effective curves
$E \subset S$.
Now that we are on a surface, the only curve $A$ which could
conceivably intersect $E$ negatively is $E$ itself; so the
condition is equivalent to $E^2<0$.

 Let $E_i,\  i=1, \ldots ,9$ be the exceptional divisors obtained by
blowing up the 9 points $p_i$, and let $H$ be (the pullback of) the
hyperplane class in $\bP^2$.  The integer cohomology of $S$ is then
$$I:=H^2(S,{\bZ})={\bZ}\cdot H+\sum^9_{i=1}{\bZ }\cdot E_i.$$
The anticanonical class is
$$F:=-K_S=3H-\sum^9_{i=1}E_i.$$

Consider first the special case where the 9 points are a complete
intersection, so there is an elliptic fibration
$$p: S\rightarrow \bP ^1$$
with fibers of cohomology class $F$.  In this case we claim that
the non-nef divisors $E$ in $S$ are precisely the sections of the
map $p$.  Indeed, adjunction gives
$$E^2=2 g-2-K_S\cdot E=2 g-2+F\cdot E.$$
This can be negative only if $g=0$ and $F\cdot E=0$ or 1, since $F$
is nef.  Now $E\cdot F=0$ is possible only if $E$ is a component of
one of the cubics $F$ through the $p_i$; but our mild general
position assumptions on $p_i$ (no three colinear, no 6 on a conic)
suffice to exclude reducible fibers. So the non-nef curves must be
$ (-1)$-curves  (smoothly embedded rational curves $E$ with $E^2=-1$)
which intersect
each $F$ in a single point, i.e. are sections of $p$.
Conversely, any section $E$ is rational, so by adjunction $E^2=-1$.

     The elliptic fibration disappears when we go to a general $S$,
but the set of classes in $I$ of the
$(-1)$-curves remains unaltered.  In
fact, an easy deformation argument shows that as $S$ varies,
its $(-1)$-curves move along with it, so the special and general
 surfaces contain
the ``same" curves.  From now on we will work exclusively with the
elliptically fibered surfaces.

 In the next section we will
identify explicitly the image in $I$ of the lattice of sections,
and thereby obtain a formula for the superpotential.

\medskip

\S 5. \bf The Superpotential. \rm

\medskip

   To compute the superpotential, we need to know all smooth
vertical divisors $D=\pi^{-1}(C)$ in $X$ with $\chi( \O _D)=1$.  In the
previous section we saw that it suffices to consider those $C$ of
the form $E\times \bP^1$, where $E$ is a section of
$p:S\rightarrow \bP^1$.  Conversely, each of these divisors does make
a non-zero contribution to the superpotential.  To see this, we use
the second elliptic fibration $\tilde{\pi}:\; X\rightarrow
\tilde{B}$ which was discussed at the end of section 3.  Since
$\tilde{\pi}$ is a fiber product of $\kappa$ and $p$, each section $E$
of $p$ gives a divisor $D=\pi^{-1}(E\times \bP^1)$ which is a
section of $\tilde{\pi}$.  In particular, $D$ is isomorphic to
$\tilde{B}$, which is the blowup of $\bP^1 \times \bP^2$ along a
curve $\Gamma$.  It follows that the Hodge numbers of $D$ vanish:
$h_1=h_2=h_3=0$. So we are precisely in the situation where the
contribution to the superpotential is known to be non-zero.

     Furthermore, the terms in the superpotential arising from all
sections have the same coefficient.  This is because for any two
sections $E,E'$ of $p$ there is an automorphism $\alpha$ of $S$
which is the identity on $\bP^1$ (i.e. it acts fiberwise) and which
sends $E$ to $E'$.  The action of $\alpha$ on each fiber is
translation (using the group structure on the fibers) by the point
corresponding to the difference $E'-E$.

    To be more precise,
the superpotential actually depends on variables (such
as the complex structure of the four-fold $X$ that is elliptically
fibered over $B$)  which we will not be including explicitly in the 
computation that follows.  The argument above shows that the contribution
to the superpotential has a dependence on the suppressed variables
which is the same for all divisors.\footnote{One of the variables
that we are suppressing perhaps deserves special mention.
While $H^2(S)$ is 10 dimensional, $H^2(B)$, which is relevant because
we are really compactifying the Type IIB superstring on $B=\bP^1\times S$,
is eleven-dimensional.  There are thus eleven Kahler modes, of which
the superpotential depends on only ten.   It can be shown that
the natural coordinates on the Kahler moduli space are the volumes
of divisors in $B$.  One can pick eleven independent divisors in $B$
as follows: ten are of the form $\bP^1\times E$ with $E$ a divisor in $S$,
and the eleventh is $S$ itself (times a point in $\bP^1$).  The variable
we are suppressing is the volume of $S$; the ten variables we keep
are invariant under multiplying the Kahler class  of $\bP^1$ by a constant
$t$ while multiplying that of $S$ by $t^{-1}$.  In suppressing the
eleventh variable, we are in essence scaling the area of $\bP^1$ to 1
and identifying the volume of $\bP^1\times E$ with the area of $E$.}

     Let $L'$ be the set of $(-1)$-curves in $S$, i.e. sections of
$p$, and $$c:\; L'\rightarrow I=H^2(S,{\bZ})$$
the map sending a curve to its cohomology class.  Up to a function
of the suppressed variables, the superpotential
is given by $$S(z)=\sum_{E \in L'}e^{2\pi i<c(E),z>}.$$
Here $z$ is a vector of 10 complex coordinates.
  Its most natural
interpretation is as an element of
$$H_2(S,{\bC})={\rm Hom}(I,{\bC}).$$

   Using the group law on the fibers, $L'$ has the structure of an
(affine) lattice.  Its lattice of translations can be thought of as
the group $L$ of sections of the relative elliptic curve (i.e.
elliptic surface with distinguished section) $Pic^0 (S / \bP^1)$.
The choice of a section, e.g. $E_1$, determines an isomorphism
$L\tilde{\rightarrow} L'$.

   Let $L_0,L'_0$ be the sublattices of $L, L'$ generated by the 9
exceptional classes $E_i,\; i=1,...,9$.  More precisely,
$$L_0=\{ \sum^9_{i=1}n_iE_i \; | \; n_i\in {\bZ},  \sum n_i=0\}$$
$$L'_0=\{ \sum^9_{i=1}n_iE_i \; | \; n_i\in {\bZ},  \sum n_i=1\},$$
where the summation is with respect to the group-law on the fibers.
We will see momentarily that these are sublattices of index 3.

     We claim that $c: L'\rightarrow I$ is injective, and its
image is precisely the set of $(-1)$-classes:

$I^{(-1)} :=  \{ c(E) \in I | \ E^2 =-1, E \cdot F =1\}$

$\phantom{I^{(-1)}}=\{d_0 H +\sum ^9 _{i=1} d_i E_i | \ d_i \in {\bZ},
 \ \sum ^9 _{i=1} d_i= 1-3 d_0, \ d_0 ^2 -
\sum ^9 _{i=1} d^2_i = -1 \}$

$\phantom{I^{(-1)}}=
\{c_0 F +  \sum ^9 _{i=1} c_i E_i | \ c_i -c_0 \in {\bZ}, \
3c_0 \in {\bZ} , \ \sum ^9 _{i=1} c_i =1, \
 c_0 = \frac{1}{2} ( \sum ^9 _{i=1} c^2_i -1). \}$

\noindent Here we set $d_0=3c_0,\; d_i=c_i-c_0\; \; (i=1,...,9)$, and use the
relations:

$$F= 3H - \sum E_i, $$
$$H^2 =1,\  F^2 =0, \ H \cdot F =3 $$
$$H \cdot E_i =0, \ F \cdot E_i =1, \ { E_i}^2= -1, \ E_i \cdot E_j =0 \;
(i\neq j).$$

 The proof that $ c$ gives an isomorphism
of $L'$ with $I^{(-1)}$ will proceed in several
stages.   The inclusion $c(L')\subset I^{(-1)}$ is clear. To see
that $c$ is injective, note that if 
$c(E)=c(E')$ then
$E'\cdot E=E\cdot E=-1$.  $E'\cdot E<0$ implies that $E$ and $E'$
have a component in common, but since the $-1$ curves, being sections, are
all irreducible, this implies $E=E'$.  So $c$ is injective.

What remains, therefore, is to show that $c$ maps onto all of $I^{(-1)}$.
We will
do this by first showing a surjection  
$c:L'_1\tilde{\rightarrow}I^{(-1)}$, where $L'_1$ is the
affine sublattice of $L'$ obtained from $L' _0$
 by adjoining a point of order 3 
mod $L'_0$. 
 Using the injectivity  $c: L' \subset I^{(-1)} $, we will conclude
that  actually $L_1'=L'$.

We start by giving $c$ explicitly on $L'_0$: 
 if $E=\sum^9_{i=1}\;
n_iE_i$ with $n_i\in {\bZ},\; \sum n_i=1$ (summation here is with
respect to the group law), then $c(E)=c_0F+\sum^9_{i=1}c_iE_i$ with
$$ c_i = n_i, \ i=1, \cdots 9, $$
$$c_0 = \frac{1}{2} ( \sum c^2 _i -1).$$
The last formula follows from the first  since we know $c(E)\in
I^{(-1)}$ (equivalently, since $E^2=-1$).  The easiest way to see
that $c_i=n_i$ is to note
\footnote{We are grateful to Ron Livne for
suggesting this simplification.}
that addition of sections (with respect to the group law)
is compatible with, and in fact is defined in terms of, linear
equivalence on the generic fiber.  Therefore, the cohomology
classes $c_i, \ ( i \geq 1)$, must be linear expressions
 in the coefficients $n_i$,
and these expressions are determined uniquely by their values on
the 9 exceptional divisors $E_i$.
     These formulas show that $c$ gives a bijection from $L'_0$ to
the subset $I^{(-1)}_0 \subset I$ where all the $c_i$ are integers. 
Since $I$ has no torsion, the same formulas must hold for any
section $E$ which equals 
(in the Neron-Severi group of the generic fiber) a
combination $\sum n_iE_i$ with rational coefficients $n_i$.  

To get all of $I^{(-1)}$, we need to find a section $E_0$ such
that $c(E)$ gives non-integer $c_i$.  To find such an $E_0$, let
$M$ be the line in $\bP^2$ between $p_2$ and $p_3$, two of the
points that are blown up to make $S$, and let $E_0$ be the proper
transform of $M$ in $S$.  As $E_0$ is a $-1$ curve, it is necessarily
a section of the elliptic fibration $S$ and hence an element of
$L'$.  Now, let us determine $c(E_0)$.  By working out the intersection
numbers of $E_0$ with the pullback of a generic line in $\bP^2$
and with the exceptional curves, one gets that the cohomology
class of $E_0$ is $c(E_0)=H-E_2-E_3$ or equivalently
$c(E_0)=(F/3)+(1/3)\sum_{i=1}^9E_i-E_2-E_3$.

 $L_1$ is then defined to be the lattice generated by $L_0$
and the class $E_0 - E_1$ for the above $E_0$.  We have
$$L_0\subset L_1 \subset L$$
and $L_0$ has index 3 in $L_1$.  Since the restriction
on integrality of the $c_i$ has been removed by adjoining $E_0-E_1$,
our formulas give a surjection
$c:L'_1\tilde{\rightarrow}I^{(-1)}$, where $L'_1$ is the affine
lattice
$$L'_1:=\{E_1+E|E\in L_1\}$$ corresponding to $L_1$.
 Using the injectivity  $c: L' \subset I^{(-1)} $, we conclude
that $L_1=L$, $L_1'=L'$, and 
$c$ is a bijection from $L'$ to $I^{(-1)}$   as claimed.

Presently, we will compute the superpotential and
exhibit its relation to $E_8$.  To make this less mysterious,
note that $L_0$ can be thought of as the root lattice of the
group $SU(9)$.  The $SU(9)$ root lattice, which has discriminant
nine, is of index three in the self-dual  root lattice
of $E_8$.  Adjoining $E$ to $L_0'$ is analogous to adjoining
(say) $E-E_1$ to $L_0$.  The lattice obtained by extending
$L_0$ in that way is  the $E_8$ lattice.  Indeed, the $E_8$
lattice can be built from the $SU(9)$ root lattice by adjoining
any weight of the third rank antisymmetric tensor representation
of $SU(9)$, or its dual.
But $E-E_1= {1\over 3}\sum_{i=1}^9E_i-E_1-E_2-E_3$ is
such a weight.

Now to  get an explicit formula for the superpotential, we choose
the following basis for $L$:

$$A_i = E_i - E_{i+1}, \ i =1, \cdots, 7$$
$$A_8 = \frac{1}{3} \sum ^9 _{i=1} E_i - (E_1 + E_2 + E_3).$$

This gives an isomorphism from ${\bZ}^8$ (with coordinates $m_i$) to
$L$, and we translate to $L'$ by adding $E_9$:

$n_1 = \phantom{ -m_0 + } m_1- \frac{2}{3} m_8$

$n_2 = -m_1 + m_2 - \frac{2}{3} m_8$

$n_3 = -m_2 + m_3 - \frac{2}{3} m_8$

$n_4 = -m_3 + m_4 + \frac{1}{3} m_8$

$n_5 = -m_4 + m_5 + \frac{1}{3} m_8$

$n_6 = -m_5 + m_6 + \frac{1}{3} m_8$

$n_7 = -m_6 + m_7 + \frac{1}{3} m_8$

$n_8 = -m_7 \phantom{+ m_8 } + \frac{1}{3} m_8$

$n_9 = \phantom{-m_1 + m_8} + \frac{1}{3} m_8 +1.$

These formulas determine the $c_i$ for $i=1,...,9$, while for $c_0$
we find:
$$c_0=\frac{\sum^9_{i=1} n_i^2-1}{2}=\sum^8_{i=1}m_i^2-(m_1\cdot
m_2+...+m_6\cdot m_7+m_3\cdot m_8)+\frac{1}{3}m_8.$$
We note that the quadratic part is precisely the intersection form
of the $E_8$ lattice, whose appearance was explained above.
We introduce new coordinates on $I \otimes {\bC}$:

$w_i = z_i -z_{i+1}, \ i =1, \cdots 7$

$w_8=\frac{1}{3} ( -2 ( z_1 +z_2 + z_3) +( z_4 + \cdots z_9) + z_0)$

$\tau = z_0,$

\noindent so that
$$\sum^9_{i=1} n_iz_i=\sum^8_{i=1}m_iw_i+(z_9-\frac{1}{3}m_8z_0).$$

\noindent We use $\exp{x}$ to denote $e^{2 \pi i x}$. 
The superpotential becomes

\noindent $S(z) = \sum _{E \in L'} \exp{<c(E),z>}$

\noindent $\phantom{S(z)}= \sum _{m \in {\bZ}^8} \exp{
(\sum ^9_{i=1} n_iE_i + c_0 z_0)}$

\noindent $\phantom{S(z)}= \sum _{m \in {\bZ}^8} \exp{
(\sum ^8_{i=1} m_iw_i  +(z_9-\frac{1}{3}m_8\tau)}$

\noindent $\phantom{S(z)=\sum _{m \in {\bZ}^8} \exp{}}
 +(\sum_{i=1}^{8} m_i^2 -
(m_1m_2+...+m_6m_7+m_3m_8) + (1/3)m_8) \tau)$

\noindent $\phantom{S(z)}=e^{ 2 \pi i z_9} \cdot
 \theta_{E_8} (\tau E_8; \ w_1, \ldots , w_8).$

Apart from the exponential prefactor,
this is the ordinary theta function of the 8-dimensional
principally polarized abelian variety with period matrix $\tau
E_8$, where $E_8$ here stands for (negative) the Cartan matrix of the
$E_8$ algebra.  We recall that this function appears in $E_8$ current
algebra; the character of the $\widehat E_8$ affine Lie algebra
at level one is $\theta_{E_8}(\tau ;w_1,\dots, w_8)/\eta(\tau)^8$,
where $\eta$ is the Dirichlet eta function.  It is an interesting
challenge to understand why $E_8$ current algebra would appear here.

\medskip

\S 6. \bf The Moduli Space. \rm

\medskip

Suppressing the  variables mentioned in the beginning of 
section 5 that do not appear in the superpotential, this
theory can be described in terms of
 a ten dimensional moduli space. It is a 
$\bf C^*$ bundle, in the $z_9$ direction, over a nine dimensional 
space $M$. This $M$ is fibered over the upper half plane (with 
coordinate $\tau = z_0$), and the fiber over $\tau$ is an eight 
dimensional abelian variety $A_\tau$, which can be described in 
several ways:

\noindent (1) $A_\tau = L \otimes E_\tau$, where $L$ is the $E_8$ 
lattice, and $E_\tau$ is the elliptic curve with modulus $\tau$.

\noindent (2) $A_\tau$ is the principally polarized abelian variety 
determined by the $8 \times 8$ period matrix $\tau E_8$.

\noindent (3) $A_\tau$ is the moduli space of degree 0 holomorphic 
$T_{\bf C}$ bundles over $E_\tau$. Here $T_{\bf C}$ is the 
complexification of the maximal torus $T$ of $E_8$.

\noindent (4) $A_\tau$ is the moduli space of flat (unitary) $T$ 
connections on $E_\tau$.

$M$ has $SL(2,{\bf Z})$ (acting on $\tau$ and $A_\tau$) 
and the $E_8$ Weyl group (acting on $A_\tau$) as  symmetries, and it is
very plausible that one should divide by these to define
the physical moduli space.

Once one has computed the superpotential
of a supersymmetric system, one of the most interesting questions
is to look for supersymmetric vacuum states, which correspond to
solutions of $W=dW=0$.  These can be interpreted as singularities
of the hypersurface defined by $W=0$.  In our case,
since the exponential $e^{2\pi i z_9}$ has no zeroes,
we can reduce the discussion to the nine-dimensional space $M$.
Supersymmetric vacua correspond to singularities of the hypersurface
in $M$ defined by $\theta_{E_8}=0$, that is solutions of $\theta_{E_8}=
d\theta_{E_8}=0$.

The following argument allows us to identify on each $A_\tau$
a Weyl orbit of 
isolated singularities or supersymmetric vacua. 
We do not know whether there are any others.
While the computation of the superpotential was based on the relation 
between the $E_8$ lattice $L=\Gamma_{E_8}$ and the $SU(9)$ lattice 
$L_0=\Gamma_{SU(9)}$, we will use
 the relation between the $E_8$ and $SO(16)$ 
lattices (which is familiar to string theorists in connection
with the fermionic construction of $E_8$ current algebra)
to find a singularity.  If ${\bf P}SO(16)=SO(16)/\{\pm 1\}$, then the 
root lattice $\Gamma_{{\bf P}SO(16)}$ has index 4 in the weight lattice
$\Gamma_{Spin(16)}$, with the character lattice $\Gamma_{SO(16)}$ as one 
of the three intermediate lattices. The other two are obtained from 
$\Gamma_{{\bf P}SO(16)}$ by adjoining the highest weight of either of 
the half-spinor representations. Either one of these is isomorphic to 
our $L=\Gamma_{E_8}$.

The theta function associated to a lattice $\Gamma$ with positive 
definite quadratic form 
$Q: \Gamma \rightarrow {\bf Z}$ is:
$$ \theta_{\Gamma}(\tau,w) := \sum_{n \in \Gamma}
                             \exp((\tau /2) Q(n) + (n,w)). $$
\noindent Here $w$ is a variable in the dual vector space 
$Hom(\Gamma, {\bf C})$. 
There are standard theta function formulas which apply to our situation, 
cf. [12]. They say:
\\ \noindent {\bf (Product Formula)}
 If $(\Gamma, Q)$ is a direct sum of sublattices 
$(\Gamma_i,Q_i)$, then $\theta_{\Gamma} = \Pi_i \theta_{\Gamma_i}$.
\\ \noindent {\bf (Riemann's Theta Identities)} 
  If $(\Gamma ',Q')$ is a sublattice 
of finite index in $(\Gamma,Q)$, 
with the induced quadratic form, then there are relations of the form:
$$\theta_{\Gamma '}(\tau,w)= (1/d) \sum_s \theta_{\Gamma}(\tau,w+s)  $$

$$\theta_{\Gamma}(\tau,w) = 
\sum_r c_r \theta_{\Gamma '}(\tau,w+\tau Qr).$$

\noindent Here $r$ ranges over coset representatives of $\Gamma$ modulo 
$\Gamma '$, $s$ ranges over coset representatives of 
$Hom(\Gamma ',{\bf Z})$ 
modulo $Hom(\Gamma,{\bf Z})$, $d$ is the index of $\Gamma '$ in $\Gamma$, 
and the $c_r$ are nonzero. (The precise value in general is 
$c_r = \exp((\tau /2)Q(r) + (r,w))$.)

These formulas apply directly to our situation. 
For a Lie group $G$, we will write $\theta_G$ for the theta
function of the weight lattice of $G$.
Since the lattice 
$\Gamma_{SO(16)}$ is ${\bf Z}^8$ with its standard quadratic form, the 
product formula gives, for $w=(w_1,...w_8)$:
$$\theta_{{SO(16)}}(\tau,w) = \Pi_{i=1}^8 \theta(\tau,w_i)$$
where $\theta(\tau,w)$ is a standard theta function of a one-dimensional
lattice, with a simple zero at $w$ a certain point of order two
and no other zeroes.
Now Riemann's theta identities express $\theta _{{E_8}}$ as the 
sum of two $\theta_{{{\bf P}SO(16)}}$ terms, each of which, in turn, 
is the sum of two $\theta_{{SO(16)}}$ terms. The result is an 
expression of the form:
$$\theta_{{E_8}}(\tau,w)=\sum_{a=1}^4 c_a \Pi_{i=1}^{8} 
\theta(\tau,w_i - u_a),$$
where $u_a$ ranges over the four points of order two,
or spin structures, of the elliptic curve
$E_{\tau}$ and $c_a$ is non-zero.
This is simply the familiar formula expressing the character of
$E_8$ current algebra in terms of free fermions, summed over spin
structures.

This expression makes it possible to exhibit, 
for each $\tau$,
a Weyl orbit of singularities of the hypersurface $\theta_{E_8}=0$.
These are the 
points where, for each $a$, 
 precisely two of the $w_i$ are equal to $u_a$, 
for instance $w_1=w_2=u_1,w_3=w_4=u_2,
w_5=w_6=u_3,w_7=w_8=u_4$, or a permutation (which would act as a  Weyl
transformation).
 At such a point, each of the four summands 
in the expression for $\theta_{E_8}$ 
vanishes to second order, so $\theta_{E_8}$ vanishes
together with all its first derivatives, giving a singularity.
 It is also clear that the singularity obtained
this way is isolated in the $w_i$ directions:
the $8 \times 8$ matrix of second order partials in the 
$w$ directions is in block diagonal form, with four $2 \times 2$ blocks 
corresponding to the spin structures. Each of these, in turn, is 
invertible, since the one-variable theta function has only first 
order zeros.  So we get in this way a supersymmetric vacuum
in which the superfields corresponding to the $w_i$, $i=1,\dots,8$
get masses, while $\tau$ and $z_9$ survive as moduli.


\medskip

\S 7 \bf Compactification To Six Dimensions. \rm

\medskip

The counting of $-1$ curves in $S$ that we have carried out actually
has another possible application.  If one compactifies Type IIB
superstring theory to six dimensions on $S$, that is if one studies
$F$-theory on a Calabi-Yau 
threefold that is elliptically fibered over $S$,
then three-branes wrapped on $-1$ curves in $S$ give certain
BPS-saturated strings, and the counting of these curves has some physical
interest.  It is expected for various reasons that counting of these
objects, and perhaps additional states
that cannot be constructed in exactly this way,
should be related to $E_8$ current algebra.
A recent discussion with references is [10]; see also
section 7.4 of [11].  What we have shown is that the counting of 
$-1$ curves gives the $E_8$ theta function; the denominator
of the $E_8$ partition function  (which is a factor
$\eta(\tau)^8$, $\eta$ being the Dirichlet eta function, and
thus is independent of the $w_i$ and $z_0$) must have another
origin, perhaps from singular objects as discussed in [10].

Research of RD was supported in part by NSF grant DMS-95-03249
and a Lady Davis Fellowship from the Hebrew University,
that of AG by NSF grant DMS-94-01495, and that of EW
by NSF grant PHY-9513835.

\end{document}